# Performance Analysis of Embarassingly Parallel Application on Cluster Computer Environment : A Case Study of Virtual Screening with Autodock Vina 1.1 on Hastinapura Cluster

Muhammad H. Hilman[1], Heru Suhartanto[1] and Arry Yanuar[2]

[1]Computer Networks, Architecture, and High Performance Computing Laboratory
Faculty of Computer Science, Universitas Indonesia

[2]Department of Pharmacy, Faculty of Mathematics and Natural Sciences, Universitas Indonesia

Email: muhammad.hilman@ui.ac.id, heru@cs.ui.ac.id, arry.yanuar@gmail.com

*Abstract*—IT-based scientific research requires high computational resources. The limitaton on funding and infrastructure led the high performance computing era from supercomputer to cluster and grid computing technology. Parallel application running well on cluster computer as well as supercomputer, one of the type is embarassingly parallel application. Many scientist loves EP because it doesn't need any sophisticated technique but gives amazing performance. This paper discusses the bionformatics research that used embarassingly parallel application and show its performance on cluster computer.

## I. Introduction

The great invention on information and technology has changed the research paradigm on many field. The researchers used virtual experiment on IT to get more accurate result and to reduce the cost of experiment. However, many problems comes when many scientist try to use the computer modeling to build virtual laboratory. Sophisticated of mathematics model, limitation of funding and research infrastructure, and the most annoying problem is limitation of computational resources to get the information result as soon as posible.

### A. Need of High Performance Computing

Science and engineering problems need many computational resources to get the relevant information on research. For example the polluted concentration substance which modeled with the Differential Ordinarry Equation System

$$\frac{\partial c_s}{\partial t} = -\left(\frac{\partial u\, c_s}{\partial x} + \frac{\partial v\, c_s}{\partial y} + \frac{\partial w\, c_s}{\partial z}\right) + \frac{\partial(K_x \partial c_s/\partial x)}{\partial x} + \frac{\partial(K_y \partial c_s/\partial y)}{\partial y} + \frac{\partial(K_z \partial c_s/\partial z)}{\partial z} + E_s(\theta,t) - (k_{1s} + k_{2s})c_s(\theta,t) + R_s(c_1, \ldots, c_q) \quad (1)$$

where s = 1, ....., q; $c_s(\theta,t)$ is concentration of pollutant s at the space θ; then u(θ,t), v(θ,t), and x(θ,t) is the wind velocity along the x, y, and z axis; $E_s(\theta,t)$ states the emission on space point θ and time t for pollutant s; $k_{1s}$ and $k_{2s}$ are coefficient of deposition dry and wet; K(θ,t) states difusion coefficient along three coordinate axis; and $R_s$ states the chemical reaction related to component s [1,2].

This mathematics model has been researched by Denmark Pollution Laboratory using q = 29. If axis grid used is x = 32, y = 32, and z = 9, it results 267,264 equations that have to be solved in each integration step along time scale to know the variation each month to get actual information of pollution. This sophisticated equation can only be solved by computers that have very large computational resources [1].

### B. Cost and Infrastructure Limitation

High computational resource that can solve the sophisticated problems like the differential ordiner equations above can only be received by supercomputer infrastructure (ex: Cray X-MP, CDC, Illiac-IV) some decades ago. Limitation of integrated circuit (IC) and processor development; and growing research on networks and protocols has been stimulating the new technology of high performance computer. The new one is cluster and grid computing technology [3].

Actually, the concepts aren't literally new. When IT researcher try to develop high performance computer in the first place, they already thought the possibility of cluster and grid based computer but at that time the protocols and networks security isn't advance as now on. So they began their research on supercomputer technology that specifically building special computer with special compilers that able to run parallel application to solve any problems that need high performance coputational roseources [4].

The development cost actually got significant aspect on developing those technologies. Supercomputer infrastructures are quite expensive to build but cluster and grid technology isn't that high in comparison. Building cluster and grid computers relatively cheap. It has motivated many researchers not only in the third world country but also in the whole world trying to implement cluster and grid computer to help their scientific computation as part of their research activity.

*C. Paper Structure*

In this paper we will discuss the performance on embarrassingly parallel application as one of the parallelization method that run on cluster computer infrastructure called 'hastinapura' in Universitas Indonesia. The paper will describe the introduction as well as the background behind research in the section 1. Section 2 will describe the hardware and software architecture of our cluster 'hastinapura'. The overview of parallel computation will be described in the section 3 as well as the term embarrassingly parallel application. Section 4 will show the result of bioinformatics case that embarrassingly parallel on cluster 'hastinapura'. The last section is discussing conclusion of our works and many future opportunities to do research on this field.

## II. HASTINAPURA CLUSTER COMPUTER ARCHITECTURE

*A. Hardware Infrastructure*

Hastinapura cluster consist of one head node, one grid portal node, one storage node and 16 worker nodes [5]. Head node is the main key of the cluster technology. All of the cluster management is done by middleware that put on this server. The head node specifications are listed below,

Table 1. Head node machine spesification

| Head Node Machine Spesification | |
|---|---|
| Machine Type | Sun Fire X2100 |
| Processor | AMD Opteron 2.2 GHz |
| Memory | 2 GB |
| Operating System | GNU/Linux Debian 3.1 |

Grid portal node on hastinapura cluster used for communication between hastinapura cluster that located on Universitas Indonesia and the other cluster on another universities. This network will established the RI-Grid Infrastructure that backed up by INHERENT, the network facilities provided by ministry of education for Indonesian higher education institution. Somehow, this infrastructure still haven't achieved yet this year. The grid portal node spesification are listed below

Table 2. Grid portal node machine spesification

| Grid Portal Node Machine Spesification | |
|---|---|
| Machine Type | Sun Fire X2100 |
| Processor | AMD Opteron 2.2 GHz |
| Memory | 2 GB |
| Operating System | GNU/Linux Debian 3.1 |

Hastinapura has 16 worker nodes that provide computational resource for the users. With dual core processors, hastinapura equal to the single system that have 32 processors. The worker nodes spesification are listed below

Table 3. Worker nodes machine spesification

| Grid Portal Node Machine Spesification | |
|---|---|
| Machine Type | Sun Fire X2100 |
| Processor | AMD Opteron 2.2 GHz |
| Memory | 1 GB |
| Operating System | GNU/Linux Debian 3.1 |

The last machine that also held an important role in the system is storage node. The storage node spesification are listed below

Table 4. Worker nodes machine spesification

| Storage Node Machine Spesification | |
|---|---|
| Machine Type | Intel PC |
| Processor | Dual Intel Xeon 2.8 GHz (HT) |
| Memory | 2 GB |
| Hard Disk | 3 x 320 GB |
| Operating System | GNU/Linux Debian 4.2 |

*B. Software Infrastructure*

Hastinapura cluster uses Globus Toolkit as middleware that handle all the software management above the machine. Globus Toolkit can manage different cluster with different machine to one grid system. Above Globus Toolkit, hastinapura uses Sun Grid Engine as a job submission and cluster resources management. MPICH is provided in hastinapura to support the MPI-based parallel application that run on the cluster [5].

Many applications has been deployed on hastinapura cluster to support research activities in Universitas Indonesia. Some of the applications are mpiBlast, sequence allignment in bioinformatics that based on MPI; GROMACS, molecular dynamic simulation software; and MPI-POV-Ray, optical modelling for object-ray interaction [6].

## III. EMBARASSINGLY PARALLEL PARADIGM

There are many paradigms to do parallel programming. One of the most favourite paradigm based on its easyness and less sophisticated control is message passing paradigm. Message passing requires the programmer to handle parallel aspect of the codes when doing the programming.

The message passing paradigm generally divided into three kind of programs [7]. First, embarassingly parallel paradigm. It has virtually no communication required; easily load balanced; and it has perfect speedup. Second, regular and synchronous. It easily statically load balanced; expect good speedup for non-local comunication; and expect reasonable speedupfor non-local communication. Third, irregular and/or asynchronous. It is difficult to load balance; communication overhead usually high; and usually can't be done efficiently using data parallel programming.

### A. EP Problems

Embarassingly parallel problems has unique criteria that can easily being distinguishad. Many real world problem can be define into EP problems. Here are the EP problems criteria [7]

1. Each element of an array can be processed independently of the others.
2. No communication required, except to combine the final result.
3. Static load balancing is usually trivial – can use any kind of distribution since communication is not a factor.
4. Dynamic load balancing can be done using a task form approach.
5. Expect perfect speedup.

Here are some illustration on how the very simple EP problems modelled.

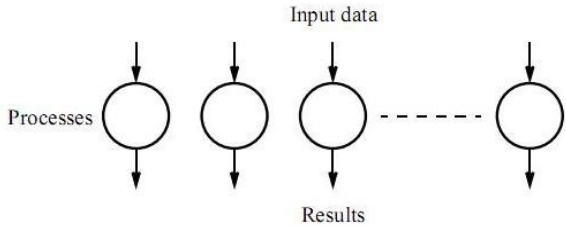

Fig 2. Disconnected computational graph [8].

More advanced EP problems based on master-slave concept of job tasking.

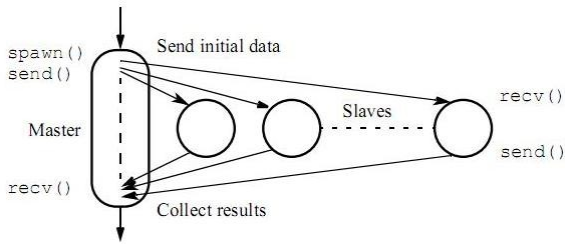

Fig 3. EP problems with dynamic process creation and the master-slave approach [8].

### B. EP Performance Measurement

Performance measurement on EP application is quite simple since no communication overhead involved. The formula for measuring the speedup is

$$speedup = \frac{time\ on\ 1\ processor}{time\ on\ N\ processors}$$

and the formula for measuring the efficiency of the EP application is

$$efficiency = \frac{speedup}{N}$$

To measure the EP application doesn't need the function that calculate some overhead involved in the process such as used in Amdahl's Law [9].

### C. EP Cases

There are many sample cases that already defined. Here are some sample on EP cases [8]

1. Geometrical Transformations of Images.
2. Mandelbrot set.
3. Dynamic Task Assignment Work Pool/Processor Farms.
4. Monte Carlo Methods.
5. Random Number Generation.

and many cases on bioinformatics that just easily perform as submitting many jobs with different parameter and different data type. In this paper we will use one of the bioinformatics case for showing the EP. The problems is molecular docking and virtual screening.

## IV. BIOINFORMATICS CASE

### A. Molecular Docking and Virtual Screening

Molecular docking is a computational procedure that attempts to predict noncovalent binding of macromolecules. The goal is to predict the bound conformations and the binding affinity [10]. The prediction process is based on information that embedded inside the chemical bond of substance.

One of the method to calculate energy that related to binding affinity aspect is AMBER (Assisted Model Building with Energy Refinement) force fields

$$V(r^N) = \sum_{bonds} \frac{1}{2} k_b (l - l_0)^2$$
$$+ \sum_{angles} \frac{1}{2} k_a (\theta - \theta_0)^2$$
$$+ \sum_{torsions} \frac{1}{2} V_n [1 + cos(\eta\omega - \gamma)]$$
$$+ \sum_{j=1}^{N-1} \sum_{i=j+1}^{N} \{\epsilon_{i,j} \left[\left(\frac{\sigma_{i,j}}{r_{i,j}}\right)^{12} - 2\left(\frac{\sigma_{i,j}}{r_{i,j}}\right)^6\right]$$
$$+ \frac{q_i q_j}{4\pi\epsilon_0 r_{i,j}}\}$$

Eq 2. AMBER force fields formula

where $\Sigma bonds$ states the distance aspect that represented by $l$ and $l_0$; $\Sigma angles$ states the angles aspect between chemical bond represented by $\theta$ and $\theta_0$; $\Sigma torsions$ states the active torsion aspect; and $\Sigma ij$ shows the non-covalent chemical bond in the substance [11].

Virtual Screening is molecular docking process that involved large database of chemical compound. We can say that virtual screening is a lot of molecular docking process to get the best molecule to become drug candidate. It's important to say that this simulation is not the only one method to get the drug candidates, practical laboratory experiment is the one to do that. This virtual experiment used to process large number of substance to produce small number of best drug candidates which cannot be done manually due large number of data. The key to parallelization process in molecular docking lies on this aspect, 'big number of molecular docking jobs'.

*B. Autodock Vina 1.1*

Autodock Vina is molecular docking software that developed by The Scripps Research Institute, non-profit biomedical research from San Diego, California, USA. Autodock Vina is the next generation of molecular docking engine after The Scripps Research Institute released Autodock in the first place.

Autodock Vina gives transparent user point of view in using molecular docking parameters. User doesn't have to do many script programming for adjusting the parameters, Autodock Vina has provided simple library that users can only put the desired parameter on docking process.

The Scripps Research Institute claimed that Autodock Vina has better performance than Autodock. This claim proved by their experimental performance that published together with the Autodock Vina software news and update in the journal of computational chemistry [10].

The most important from Autodock Vina is the application can run on multicore machine. Autodock Vina has implemented hyper threading concept that run perfectly on a machine that has more than one processor. But the hard point in this paper is not the capability of Autodock Vina running on multicore machine; we will discuss the input problem of virtual screening in Autodock Vina that has very likely type with EP problems.

*C. Experimental Result*

Here we will show the result of virtual screening experiment that uses Autodock Vina on hastinapura. The ligand database uses in this experiment was taken from ZINC (ZINC Is Not Commercial) Database in the mol2 format and the receptor was taken from PDB (Protein Data Bank) Database in the pdb format. We process both the input to the pdbqt format and begin the virtual screening process after setting some parameters on both ligands and receptor.

We do the experiment on two different system to get comparison how is the performance. First, we experiment on sequential paradigm by running it locally on PC desktop with the specification as listed below

Table 5. PC desktop machine spesification

| PC desktop Machine Spesification | |
|---|---|
| Machine Type | Intel PC |
| Processor | Pentium IV, 2.00 GHz |
| Memory | 1 GB |
| Hard Disk | 80 GB |
| Operating System | GNU/Linux Ubuntu 8.0 |

Second, we run the parallel experiment in the hastinapura cluster using qsub job submission script that used in Sun Grid Engine. The cluster machine specification is discussed in the section before.

We use different numbers of data in these two different machines to get the information about speedup and efficiency. The experimental show the result as follows

Table 6. Experimental result

| Data | Execution Time (minutes) | |
|---|---|---|
| | Serial | Parallel |
| 1000 | 2277.42 | 77.43 |
| 2000 | 4629.72 | 159.5 |
| 3000 | 8117.6 | 292.27 |
| 4000 | 12370.5 | 406.8 |
| 5000 | 15294.2 | 509.8 |

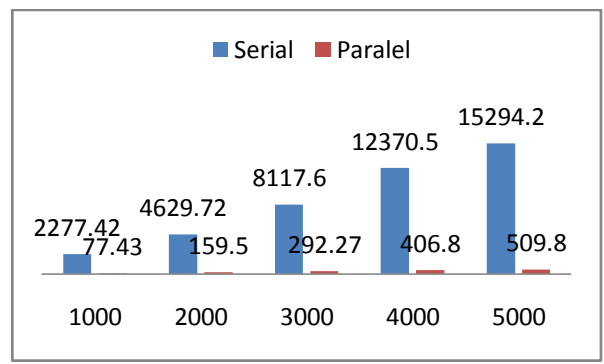

Fig 4. Experimental result (general)

From the table and diagram above we can see that running time is accelerated 29.16 times. We also can conclude that speed up on this experiment is 29.16 times because serial experiment is done on machine with one cpu.

Beside the running time and speed up, the efficiency can be calculated from information we got. Efficiency of the experiment is 91%, almost linier. We just can see it clearly from two diagrams below about the linierity of the experiment. These three diagrams and one table show clearly that performance of embarassingly parallel application is generally linier and give good result on speed up and efficiency.

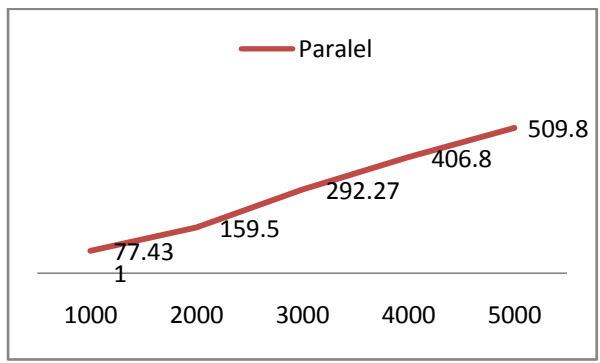

Fig 5. Experimental result (parallel)

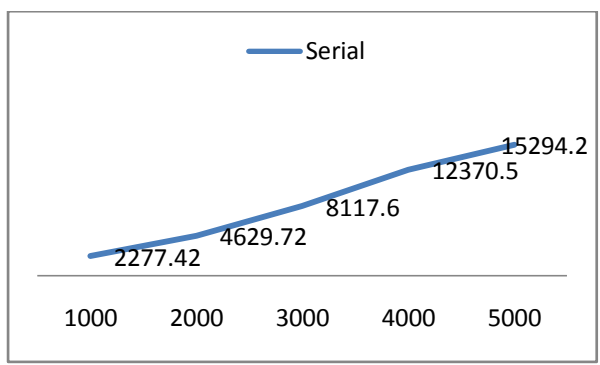

Fig 6. Experimental result (serial)

Fig 5., and fig 6. Show the linierity of virtual screening experiment that run on the hastinapura cluster computer.

## V. CONCLUSION AND FUTURE WORKS

### A. Conclusion

Scientific research with the spesific problems that can be divide into many jobs which does not have high dependencies in each proces can be define into embarassingly parallel problems. EP problems runs well on cluster computer environment that can be build as the substitution of supercomputer infrasturcture. EP has became favorite between researchers that need large computational resources. EP gives very nice result on cluster computer environment. It show good speed up and efficiency regarding general performance of application.

### B. Future Works

The open research problem on this area is concerned on the hardware of cluster infrastructure. Simple explanation on EP is the application submitting many jobs concurrently, but actually the job submission engine and the machine itself exploited too much on running this application. So, it will become big problem if there is some interruption on job submission or the machine itself. The new software and hardware configuration for cluster computer environment should be designed well to handle that problems.

ACKNOWLEDGMENT

We would like to thank directorate of research and social service of Universitas Indonesia that already gave financial support on the whole project of biomolecular and drug design computational research using hastinapura cluster computer in Universitas Indonesia.